\documentclass[aps,pra,reprint,superscriptaddress]{revtex4-2}

\usepackage{amssymb,amsmath,amsfonts,amsthm}
\usepackage{graphicx}
\usepackage{hyperref}
\usepackage{bbm}
\usepackage{color,xcolor}

\begin{document}


\title{Collective-motion-enhanced acceleration sensing via an optically levitated microsphere array}


\author{Yao Li}
\affiliation{State Key Laboratory of Modern Optical Instrumentation ${\&}$ College of Optical Science and Engineering, Zhejiang University, Hangzhou, 310027, China}

\author{Chuang Li}
\email[]{lic@zhejianglab.com}
\affiliation{Research Center for Quantum Sensing, Zhejiang Lab, Hangzhou, 311121, China}

\author{Jiandong Zhang}
\affiliation{School of Mathematics and Physics, Jiangsu University of Technology, Changzhou, 213001, China}

\author{Ying Dong}
\email[]{yingdong@zhejianglab.edu.cn}
\affiliation{Research Center for Quantum Sensing, Zhejiang Lab, Hangzhou, 311121, China}

\author{Huizhu Hu}
\affiliation{State Key Laboratory of Modern Optical Instrumentation ${\&}$ College of Optical Science and Engineering, Zhejiang University, Hangzhou, 310027, China}
\affiliation{Research Center for Quantum Sensing, Zhejiang Lab, Hangzhou, 311121, China}


\date{\today}

\begin{abstract}
Optically levitated microspheres are an excellent candidate for force and acceleration sensing.
Here, we propose an acceleration sensing protocol based on an optically levitated microsphere array (MSA).
The system consists of an $N$-microsphere array levitated in a driven optical cavity via holographic optical tweezers.
By positioning the microspheres suitably relative to the cavity, only one of the collective modes of the MSA is coupled to the cavity mode. The optomechanical interaction encodes the information of acceleration acting on the MSA onto the intracavity photons, which can then be detected directly at the output of the cavity. The optically levitated MSA forms an effective large mass-distributed particle, which not only circumvents the problem of levitating a large mass microsphere but also results in a significant improvement of sensitivity. Compared with the traditional single-microsphere measurement scheme, our method presents an improvement in sensitivity by a factor of $\sqrt{N}$.
\end{abstract}


\maketitle

\section{Introduction}


Optomechanics is a fascinating field that explores the interaction between optical and mechanical degrees of freedom, and it has a wide range of applications in various scientific domains \cite{aspelmeyer2014cavity}. One of the most promising applications of optomechanics is in precision measurement, where optomechanical sensors enable the ultra-sensitive detection of forces and accelerations.\cite{anetsberger2009near,gavartin2012hybrid,krause2012high,guzman2014high,mason2019continuous,zhao2020weak}. 

Levitated optomechanics \cite{gonzalez2021levitodynamics}, where nano- and micro-objects are levitated in vacuum by optical tweezers \cite{ashkin1971optical,ashkin1976optical,ashkin1977feedback}, has emerged as a particularly promising area of research that has gained significant attention 
\cite{ashkin1987optical,ashkin1987optical_1,fazal2011optical,dholakia2011shaping,padgett2011tweezers,marago2013optical}. Optically levitated sensors provide high isolation from thermal and environmental sources of noise in high-vacuum, enabling exceptional displacement, force, and acceleration sensitivity \cite{ranjit2015attonewton,ranjit2016zeptonewton,hempston2017force,blakemore2019three}.
Experimental results have demonstrated force sensitivity as low as $10^{-20}, \mathrm{N/\sqrt{Hz}}$ \cite{tebbenjohanns2019cold,tebbenjohanns2020motional} and acceleration sensitivity below sub-$\mathrm{\mu g/\sqrt{Hz}}$ ($\mathrm{g} = 9.8 \mathrm{m/s^2}$) \cite{monteiro2017optical,rider2018single,monteiro2020force}. Levitated optomechanical systems offer precise control of the nanoparticle's translation and rotational degrees of freedom \cite{barker2010doppler,gieseler2012subkelvin,kiesel2013cavity,millen2015cavity,hebestreit2018sensing,delic2019cavity,hoang2016torsional}, making them an ideal platform for fundamental scientific research \cite{li2010measurement,li2013brownian,gieseler2013thermal,moore2014search,gieseler2014dynamic,jain2016direct,rider2016search,rondin2017direct,nie2014generating,carney2021mechanical,moore2021searching}, such as the detection of high-frequency gravitational wave \cite{arvanitaki2013detecting,winstone2022optical}, the search for dark matter \cite{monteiro2020search}, and the exploration of macroscopic quantum superposition with mesoscopic particles \cite{romero2010toward,romero2011large,delic2020cooling,magrini2021real,tebbenjohanns2021quantum}. Additionally, electrostatic and magnetic levitation \cite{gonzalez2021levitodynamics,lewandowski2021high}, as well as coupling to solid-state spin systems \cite{hoang2016electron,gieseler2020single} have extended the capabilities of levitated optomechanics and open up new opportunities for both scientific research and technological applications.

In the fields of precision measurements and searching for new physics, detecting forces and accelerations with increasing sensitivity is crucial \cite{moore2021searching,gonzalez2021levitodynamics}.
When it comes to acceleration sensing, a large mass nanoparticle is advantageous because the acceleration sensitivity $\sqrt{S_{a}}$ is inversely proportional to its mass $m$ \cite{millen2020optomechanics}, i.e.,
\begin{equation}\label{eq:sensitivity_single}
	\sqrt{S_{a}} = \sqrt{\frac{2 k_{\mathrm{B}}T\gamma}{m}},
\end{equation}
where $\gamma$ is the damping rate, $k_{\mathrm{B}}$ the Boltzmann constant, and $T$ the temperature of the surrounding environment. However, optically levitating a large mass nanoparticle experimentally is a huge challenge because laser heating constrains the maximum size of particles that can be levitated \cite{monteiro2017optical}. 
Another approach to improve sensitivity is by reducing the pressure in the surrounding environment. Levitated optomechanical experiments have been conducted at a pressure level of around $10^{-9}$ mbar \cite{magrini2021real,tebbenjohanns2021quantum}. However, a further reduction in pressure is still a formidable challenge.



Here, we present a novel acceleration sensing method based on an optically levitated MSA.
By employing the holographic optical tweezers \cite{yang2021optical} which can generate multiple controllable optical traps, an array of small mass microspheres is levitated in vacuum forming an effective large mass sphere.
The mass of the effective sphere is distributed in multiple optical traps, thus circumventing the problem of the maximum size of spheres that can be levitated in a single optical trap.
Moreover, the effective large mass sphere will lead to a significant improvement in acceleration sensitivity when its motion of the center of mass (COM) is used for acceleration sensing.
For reading out the motion of the MSA that contains the acceleration information, we introduce a driven Fabry-Perot (FP) cavity that couples with all the microspheres.
The optomechanical interaction between the cavity field and microspheres imprints the collective motion of the spheres  on the optical field, and thus the acceleration information can be obtained by measuring the cavity field.

\begin{figure}[htbp]
	\centering
	\includegraphics[width=0.45\textwidth]{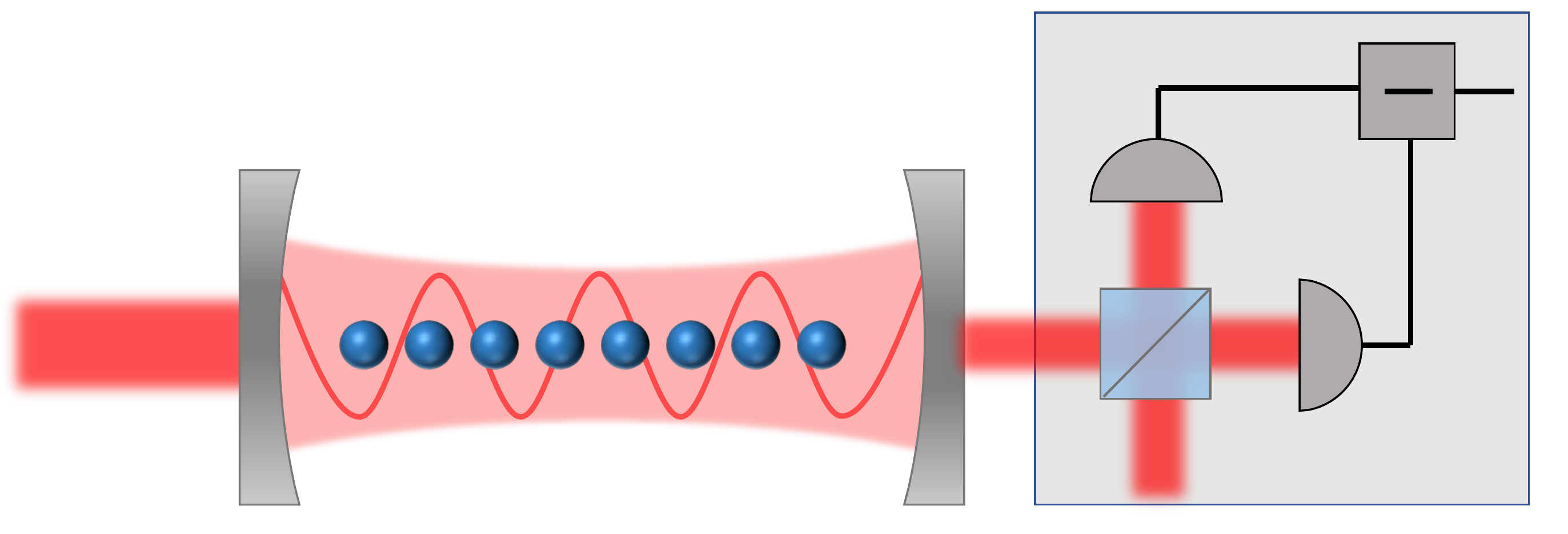}
	\caption{A simplified schematic of the experimental setup. The system consists of a $N$-microsphere array levitated in an FP cavity by a holographic optical tweezer. The FP cavity is driven by a laser and the MSA is coupled to the  cavity mode. The collective motion of the MSA responding to the external acceleration is monitored by detecting the photons leaking from the cavity.}
	\label{fig:setup}
\end{figure}

A simplified schematic of the experimental setup is shown in Fig. \ref{fig:setup}, where an MSA is trapped and levitated in an FP cavity by a holographic optical tweezer. An external laser drives the cavity and excites a standing wave mode. By placing the microspheres at suitable positions relative to the cavity, the cavity mode only couples to one of the collective modes of the MSA which takes the information of the acceleration acting on the MSA.
Then by measuring the correlation function of photons leaking from the cavity via optical methods such as homodyne or heterodyne detection at the output, we can acquire the acceleration information.

\section{\label{s2} The scheme of acceleration sensing}
\textit{The physical model} - As introduced above, we consider an optically levitated MSA where $N$ dielectric microspheres with mass $m_i$ respectively are trapped and levitated in an FP cavity which is driven by a laser, forming a multi-oscillator cavity optomechanical system (see Fig. \ref{fig:setup}).
The Hamiltonian of the whole system in a frame rotating at pumping frequency $\nu_p$ reads
\begin{equation}\label{hamiltonian}
	\begin{aligned}
		H =& \hbar\Delta_c\hat{a}^\dagger\hat{a} + \hbar\epsilon\left(\hat{a} + \hat{a}^{\dagger}\right)
		\\
		&+ \sum_{i=1}^{N}\left[\frac{\hbar\Omega_i}{2}\left(\hat{p}_i^2 + \hat{q}_i^2\right)
		+ \hbar f_{i}\hat{q}_i + \hat{H}_i^{\mathrm{Int}}\right],
	\end{aligned}
\end{equation}
where $\hat{a}$ is the annihilation operator of the cavity mode,  $\Delta_c = \omega_{c}-\nu_p$ is the detuning of the cavity mode with frequency $ \omega_{c}$ from the pumping laser, and $\epsilon$ is the strength of the pump.
$\hat{q}_{i}$ and $\hat{p}_{i}$ denote respectively the dimensionless position and momentum operators of the mechanical oscillators, and $\Omega_{i}$ is the frequencies of the mechanical oscillators. $f_i$ is the external force acting on each mechanical oscillator, which induces the inertial acceleration $a_i$ via
\begin{equation}\label{eq:force_to_acceleration}
	f_i = \sqrt{\frac{m_{i}}{\hbar \Omega_{i}}} a_i.
\end{equation}
$\hat{H}_{i}^{\mathrm{Int}}$ accounts for the linear coupling between the cavity mode and the mechanical oscillators. In the case that the radii of spheres are much smaller than the cavity mode wavelength $r_{i} \ll \lambda_{c}$, the dipole force potential created by the cavity mode can be approximated as \cite{chang2010}
\begin{equation}
	\begin{aligned}
		\hat{U}_{i}^{\mathrm{dipole}} = & - \hbar g_i^0\hat{a}^{\dagger}\hat{a}\cos\left[2k(q_{i}^{\mathrm{eq}} + \hat{q}_{i}) \right],
	\end{aligned}
\end{equation}
where $k = 2\pi/\lambda_c$ denotes the wave number of cavity field, $q_{i}^{\mathrm{eq}}$ is the equilibrium position of the $i$th oscillator, and $g_i^0 = \frac{3 V_i}{4 V_{c}} \frac{\epsilon_0 - 1}{\epsilon_0 + 2}\omega_{c}$ is the single photon optomechanical coupling strength with $V_i$ and $V_{c}$ the volumes of the mircospheres and the cavity respectively and $\epsilon_0$ the electric permittivity. Assuming the oscillation of the sphere around its equilibrium position is small, one can expand  the dipole force potential to first order and obtain a linear-coupling interaction Hamiltonian
\begin{equation}
	\hat{H}_{i}^{\mathrm{Int}} = \hbar g_{i}\hat{a}^{\dagger}\hat{a}\hat{q}_{i},
\end{equation}
where $g_{i} = -2g_i^0 k \sin [2 k q_{i}^{\mathrm{eq}}]$ is the linear-coupling strength, whose value can be adjusted by shifting the equilibrium positions $q_{i}^{\mathrm{eq}}$.

\textit{The dynamical equations and solutions} - In strong driving regime, we can rewrite each operator as the sum of its classical steady-state mean value and a small quantum fluctuation operator, i.e., $\hat{a} \rightarrow \alpha + \hat{a}$, $\hat{q}_i \rightarrow q_i^{s} + \hat{q}_i$ and $\hat{p}_i \rightarrow p_i^{s} + \hat{p}_i$. Employing the standard linearization procedure \cite{genes2008ground}, we obtain the quantum Langevin equations (QLEs) in terms of the quantum fluctuating operators 
\begin{eqnarray}\label{QLEs}
	\dot{\hat{a}}\ &=& -\left(\mathrm{i}\Delta+ \frac{\kappa}{2}\right)\hat{a}
	- \sum_{i}^{N}\mathrm{i}G_i\hat{q}_i + \sqrt{\kappa}\hat{a}^{\mathrm{in}}, \nonumber \\
	\dot{\hat{q}}_i &=&\  \Omega_i\hat{p}_i,
	\\			
	\dot{\hat{p}}_i &=& - \Omega_i\hat{q}_i - \gamma_i\hat{p}_i - G_i\left(\hat{a} + \hat{a}^{\dagger} \right) - f_i
	+ \hat{\xi}_i, \nonumber
\end{eqnarray}
where we have phenomenologically introduced $\kappa$ and $\gamma_i$ accounting for the dissipation of the cavity field and the mechanical oscillators, respectively. 
At the same time, we also obtain a set of classical Langevin equations (CLEs) from which the steady-state values can be figured out (see the Appendix for details). $\Delta = \Delta_c + \sum_i g_iq_i^{\mathrm{s}} $ and $G_{i} = \alpha g_{i}$ are effective detuning and coupling strength, depending on the steady-state values  obtained from the CLEs.
The operator $\hat{a}^{\mathrm{in}}$ in Eqs. (\ref{QLEs}) accounts for the input noise of the
cavity field, which has zero mean value and the correlation function
\begin{equation}\label{eq:optical_correlation}
	\langle \hat{a}^\mathrm{in}(t)\hat{a}^{\mathrm{in}, \dagger}(t^{\prime})\rangle = \delta(t - t^{\prime}),
\end{equation}
where we have assumed the equilibrium mean photon number of the thermal bath 
$$\bar{n}^{\mathrm{c}} = \left(\exp \left\{\hbar \omega_{c} / k_{B} T\right\}-1\right)^{-1} \simeq 0$$
due to the high optical frequency $\hbar \omega_{\mathrm{c}} / k_{B} T \gg 1$.
$\hat{\xi}_{i}$ is the input noise of the Brownian force acting on the $i$th mechanical oscillator, whose correlation function is given by
\begin{equation} 
	\begin{aligned}
		\langle \hat{\xi}_i(t)\hat{\xi}_j(t^{\prime})\rangle =&
		\frac{\gamma_i}{\Omega_i} \int \frac{\mathrm{d} \omega}{2 \pi} e^{-\mathrm{i} \omega\left(t-t^{\prime}\right)} \omega
		\\
		&\times
		\left[\coth\left(\frac{\hbar \omega}{2 k_{B} T}\right)+1\right]\delta_{ij},
	\end{aligned}
\end{equation}
with $\delta_{ij}$ the Kroneker symbol. The involved mechanical frequencies are never larger than hundreds of MHz and therefore one can make the approximation $\frac{\gamma_i}{\Omega_i}\omega\coth\left(\frac{\hbar \omega}{2 k_{B} T}\right) \simeq \gamma_i\frac{2 k_{B} T}{\hbar \Omega_i} \simeq \gamma_i\left(2\bar{n}^{\mathrm{m}}_i + 1\right)$ with $\bar{n}^{\mathrm{m}}_i = \left(\exp \left\{\hbar \Omega_i / k_{B} T\right\} - 1\right)^{-1}$ the mean thermal phonon number. As a result, the correlation function of Brownian noise can be safely considered Markovian \cite{genes2008ground,genes2008simultaneous}, i.e.,
\begin{equation}\label{eq:mechanical_correlation}
	\begin{aligned}
		\langle \hat{\xi}_i(t)\hat{\xi}_j(t^{\prime})\rangle \simeq&
		\gamma_i\left(2\bar{n}^{\mathrm{m}}_i + 1\right)\delta(t - t^{\prime})\delta_{ij}.
	\end{aligned}
\end{equation}

The Eqs. (\ref{QLEs}) show that if the couplings between each oscillator and the cavity field are the same, that is $G_i = G$, only the collective mode $\hat{Q} = \sum_i^N \hat{q}_i$ couples to the cavity field. Meanwhile, considering that one can choose the microspheres with the (almost) identical size, the environments around these closely positioned microspheres ($\sim \mathrm{\mu m}$) are the same, and the mechanical resonance frequency can be precisely controlled by tuning the trapping laser power, we reasonably assume that $m_i = m$, $\gamma_i = \gamma$, and $\Omega_i = \Omega$. 
The QLEs are therefore rewritten in a closed form as 
\begin{equation}\label{eq:qles-2}
	\begin{aligned}
		&\dot{\hat{X}} = \Delta\hat{Y} - \frac{\kappa}{2}\hat{X} + \sqrt{\kappa}\hat{X}^{\mathrm{in}},
		\\
		&\dot{\hat{Y}} = - \Delta\hat{X} - \frac{\kappa}{2}\hat{Y} - \sqrt{2}G\hat{Q} + \sqrt{\kappa}\hat{Y}^{\mathrm{in}},
		\\
		&\ddot{\hat{Q}}  + \gamma\dot{\hat{Q}} + \Omega^2\hat{Q} = - \sqrt{2}N\Omega G\hat{X} - \Omega F + \Omega\hat{\xi},
	\end{aligned}
\end{equation}
where $\hat{\xi} = \sum_i^N \hat{\xi}_i$ and $F = \sum_i^N f_i$ are the total Brownian noise operator and the total external force, respectively.
Without loss of generality, we assume each mechanical oscillator undergoes the same acceleration $a_i = a$ and the total external force can be rewritten as 
\begin{equation}
	F = Nf = N \sqrt{\frac{m}{\hbar \Omega}} a.
\end{equation}
For convenience, we have introduced here the amplitude and phase quadratures of the cavity field as
\begin{equation}
	\begin{aligned}
		\hat{X} &= \frac{\hat{a} + \hat{a}^{\dagger}}{\sqrt{2}}, \quad
		\hat{Y} = \frac{\hat{a} - \hat{a}^{\dagger}}{\mathrm{i}\sqrt{2}}.
	\end{aligned}
\end{equation}
The QLEs. (\ref{eq:qles-2}) clearly show that the information of acceleration is encoded into the cavity field mediated through the collective mode $\hat{Q}$. As mentioned above, the optomechanical coupling strengths need to meet the condition $G_i = G$, which can be achieved by adjusting the trapping equilibrium positions of the microspheres so that they satisfy
\begin{equation}\label{eq:position_condition}
    q_{i + 1}^{\mathrm{eq}} - q_{i}^{\mathrm{eq}} = n\lambda_{c}/2
\end{equation}
where $i = 1, 2, \dots, N-1$ and $n$ is an integer.
In state-of-the-art optomechanical experiments with levitated nano-particles, the manipulation of the trapping particle position with a step size of $8$ nm has been achieved \cite{delic2020cooling}, which is less than $1\%$ compared to the standing wavelength. The position manipulation precision can be further improved by using better nano-positioners.

By Fourier transforming the QLEs. (\ref{eq:qles-2}) into the frequency domain and using the standard input-output relations $\hat{X}^{\mathrm{out}} = \hat{X}^{\mathrm{in}} - \sqrt{\kappa}\hat{X}$, one can solve the equations and obtain the power spectral density (PSD) of the output optical field in amplitude quadrature for instance in terms of the PSDs of acceleration and the input noise as follows
\begin{equation}\label{eq:PSD-Sxx}
	\begin{aligned}
		S_{XX}^{\mathrm{out}}(\omega) =&
		\frac{N^2 m}{\hbar\Omega}|\sqrt{\kappa} H_{XF}(\omega)|^2 S_{aa}(\omega)
		\\
		&+ |\sqrt{\kappa} H_{X\xi}(\omega)|^2 S_{\xi\xi}(\omega)
		\\
		&+|\sqrt{\kappa}H_{XY}(\omega)|^2S_{YY}^{\mathrm{in}}(\omega)
		\\
		&+ |1 - \sqrt{\kappa}H_{XX}(\omega)|^2S_{XX}^{\mathrm{in}}(\omega),
	\end{aligned}
\end{equation}
from which, one can solve eventually the PSD of the acceleration in terms of the PSDs of the output optical field in amplitude quadrature and input noises as
\begin{equation}\label{eq:PSD_aa}
	\begin{aligned}
		S_{aa}(\omega) =& \frac{\hbar\Omega}{N^2 m}\Bigg\{\frac{S_{XX}^{\mathrm{out}}(\omega)}{|\sqrt{\kappa} H_{XF}(\omega)|^2}
		- S_{\xi\xi}(\omega)
		\\
		&- \left|\frac{H_{XY}(\omega)}{H_{XF}(\omega)}\right|^2S_{YY}^{\mathrm{in}}(\omega)
		\\
		&- \left|\frac{1 - \sqrt{\kappa}H_{XX}(\omega)}{\sqrt{\kappa}H_{XF}(\omega)}\right|^2S_{XX}^{\mathrm{in}}(\omega)\Bigg\},
	\end{aligned}
\end{equation}
where the symbol $H_{AB}(\omega)$ denotes the response function (whose concrete form can be found in the Appendix) from the external noise or signal source $B$ to the detectable quadrature $A$, and the PSD for quantity $\hat{O}$ is defined as $S_{OO}(\omega) = \int_{-\infty}^{\infty}\langle \hat{O}^{\dagger}(t)\hat{O}(0)\rangle\exp(\mathrm{i}\omega t)\mathrm{d}t$. It is noteworthy that the area under the experimentally measured PSD yields the variance of the acceleration \cite{aspelmeyer2014}, i.e.,
\begin{equation}\label{PSD-2}
	\langle a^2 \rangle = \int_{-\infty}^{\infty}S_{aa}(\omega)\frac{\mathrm{d}\omega}{2\pi}.
\end{equation}
Thus one can acquire both the frequency and amplitude information of the acceleration from its PSD.
The Brownian noise can be calculated from Eq. (\ref{eq:mechanical_correlation}) as $S_{\xi_{i}\xi_{i}}\left(\omega \right) = \gamma_i(2\bar{n}^{\mathrm{m}}_{i} + 1)$ and the optical input noisy PSDs are simply $S_{XX}^{\mathrm{in}}\left(\omega \right) = S_{YY}^{\mathrm{in}}\left(\omega \right) = 1/2$ for the vacuum optical reservoir. After measuring the PSD $S_{XX}^{\mathrm{out}}(\omega)$ in amplitude quadrature (or $S_{YY}^{\mathrm{out}}(\omega)$ in phase quadrature) that can be done directly with the homodyne or heterodyne detection \cite{delic2019cavity,delic2020cooling,magrini2021real} we will eventually be able to carry out the PSD of acceleration.

\section{The improvement of acceleration sensitivity through the collective motion}
To estimate the acceleration $S_{aa}(\omega)$, we invert the measured PSD of the output optical field in Eq. (\ref{eq:PSD-Sxx}) and obtain
\begin{equation}
	\begin{aligned}
		\frac{\hbar\Omega}{N^2 m}\frac{S_{XX}^{\mathrm{out}}(\omega)}{|\sqrt{\kappa} H_{XF}(\omega)|^2} =& S_{aa}(\omega)
		+ \frac{\hbar\Omega }{N^2 m}\left[S_{\xi\xi}(\omega) + S_{\mathrm{opt}}^{\mathrm{in}}(\omega)\right],
	\end{aligned}
\end{equation}
where the optical input noise
\begin{equation}
	\begin{aligned}
		S_{\mathrm{opt}}^{\mathrm{in}}(\omega) =& \left|\frac{H_{XY}(\omega)}{H_{XF}(\omega)}\right|^2S_{YY}^{\mathrm{in}}(\omega)
		\\
		&+ \left|\frac{1 - \sqrt{\kappa}H_{XX}(\omega)}{\sqrt{\kappa}H_{XF}(\omega)}\right|^2S_{XX}^{\mathrm{in}}(\omega).
	\end{aligned}
\end{equation}
Therefore the signal-to-noise ratio ($SNR$) turns out to be 
\begin{equation}
	SNR = \sqrt{\frac{S_{aa}(\omega)}{\frac{\hbar\Omega}{N^2 m}\left[S_{\xi\xi}(\omega) + S_{\mathrm{opt}}^{\mathrm{in}}(\omega)\right]}}.
\end{equation}
By setting $SNR = 1$ as the condition that defines sensitivity, we obtain the acceleration sensitivity
as
\begin{equation}
	\sqrt{S_{a}^{(N)}(\omega)} = \sqrt{\frac{\hbar\Omega}{N^2 m}\left[S_{\xi\xi}(\omega) + S_{\mathrm{opt}}^{\mathrm{in}}(\omega)\right]}.
\end{equation}
The Brownian noise acting on each mechanical oscillator is independent and uncorrelated, thus the PSD of the total Brownian noise $S_{\xi\xi}(\omega)$ is the sum of the PSD of the Brownian noise acting on every single mechanical oscillator $S_{\xi_{i}\xi_{i}}(\omega)$, which is given by
\begin{equation}
	S_{\xi\xi}(\omega) = \sum_i^N S_{\xi_i\xi_i}(\omega) = N S_{\xi\xi}^{(1)}(\omega),
\end{equation}
where $S_{\xi\xi}^{(1)}(\omega) = \gamma(2\bar{n}^m + 1)$ is the PSD of the Brownian noise acting on a single mechanical oscillator. Therefore the acceleration sensitivity is simplified as
\begin{equation}
	\sqrt{S_{a}^{(N)}(\omega)} = \sqrt{\frac{\hbar\Omega}{m}\left[\frac{S_{\xi\xi}^{(1)}(\omega)}{N} + \frac{S_{\mathrm{opt}}^{\mathrm{in}}(\omega)}{N^2}\right]}.
\end{equation}
This equation clearly shows that the acceleration sensitivity improves in increasing the number of oscillators $N$. In the limit of $k_{\mathrm{B}}T \gg \hbar\Omega$, i.e., the dissipative dynamics of the system are dominated solely by the thermal noise, the optical input noise can be omitted. As a result, we find a $\sqrt{N}$-times improvement in  sensitivity compared with the standard single-oscillator acceleration sensing scheme (see Eq. (\ref{eq:sensitivity_single})),
\begin{equation}\label{eq:sen_2}
	\begin{aligned}
		\sqrt{S_{a}^{(N)}} &\approx \sqrt{\frac{\hbar\Omega}{Nm}\gamma\left(2\bar{n}^m + 1\right)} \approx \sqrt{\frac{S_{a}}{N}}.
	\end{aligned}
\end{equation}

\begin{figure}[htbp]
	\centering
	\includegraphics[width=0.45\textwidth]{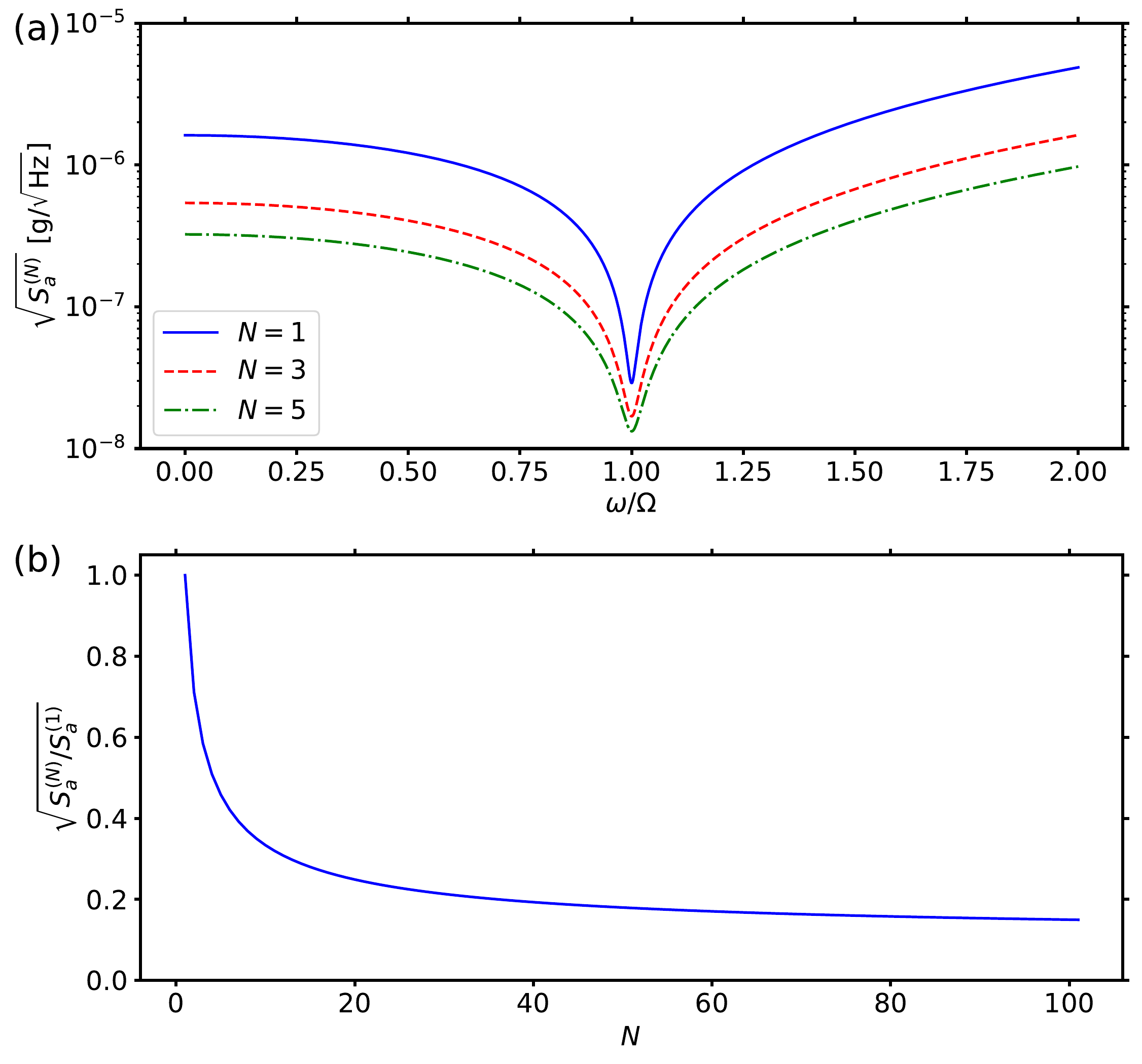}
	\caption{(a) The acceleration sensitivity as a function of frequency for different numbers of the oscillators. (b) The acceleration sensitivity at the mechanical resonance frequency (normalized with the sensitivity for $N=1$) as a function of the number of oscillators $N$. The parameters are set as $\Omega/2\pi = 100$ kHz, $\Delta/2\pi = 500$ kHz, $G/2\pi = 10$ kHz, $\kappa/2\pi = 1$ GHz, $T = 300$ K, $\gamma/2\pi = 10$ $\mathrm{\mu Hz}$ \cite{martinetz2018gas}, $r = 5$ $\mathrm{\mu m}$.}
	\label{fig:sensitivity}
\end{figure}

Figure \ref{fig:sensitivity}(a) shows the acceleration sensitivity as a function of frequency for different numbers of the oscillators $N$. The results indicate that the acceleration sensitivity improves with the increase of the oscillator number $N$. The acceleration sensitivity drops at the mechanical resonance frequency $\Omega$ due to the Lorentzian-like line shape response function. Figure \ref{fig:sensitivity}(b) shows the acceleration sensitivity at the mechanical resonance frequency (normalized with the sensitivity for $N=1$) as a function of the number of oscillators $N$. The sensitivity represents a significant improvement in increasing the oscillator number. The higher sensitivity can be achieved by using more oscillators.

\section{The numerical simulation}
\begin{figure}[htbp]
	\centering
	\includegraphics[width=0.45\textwidth]{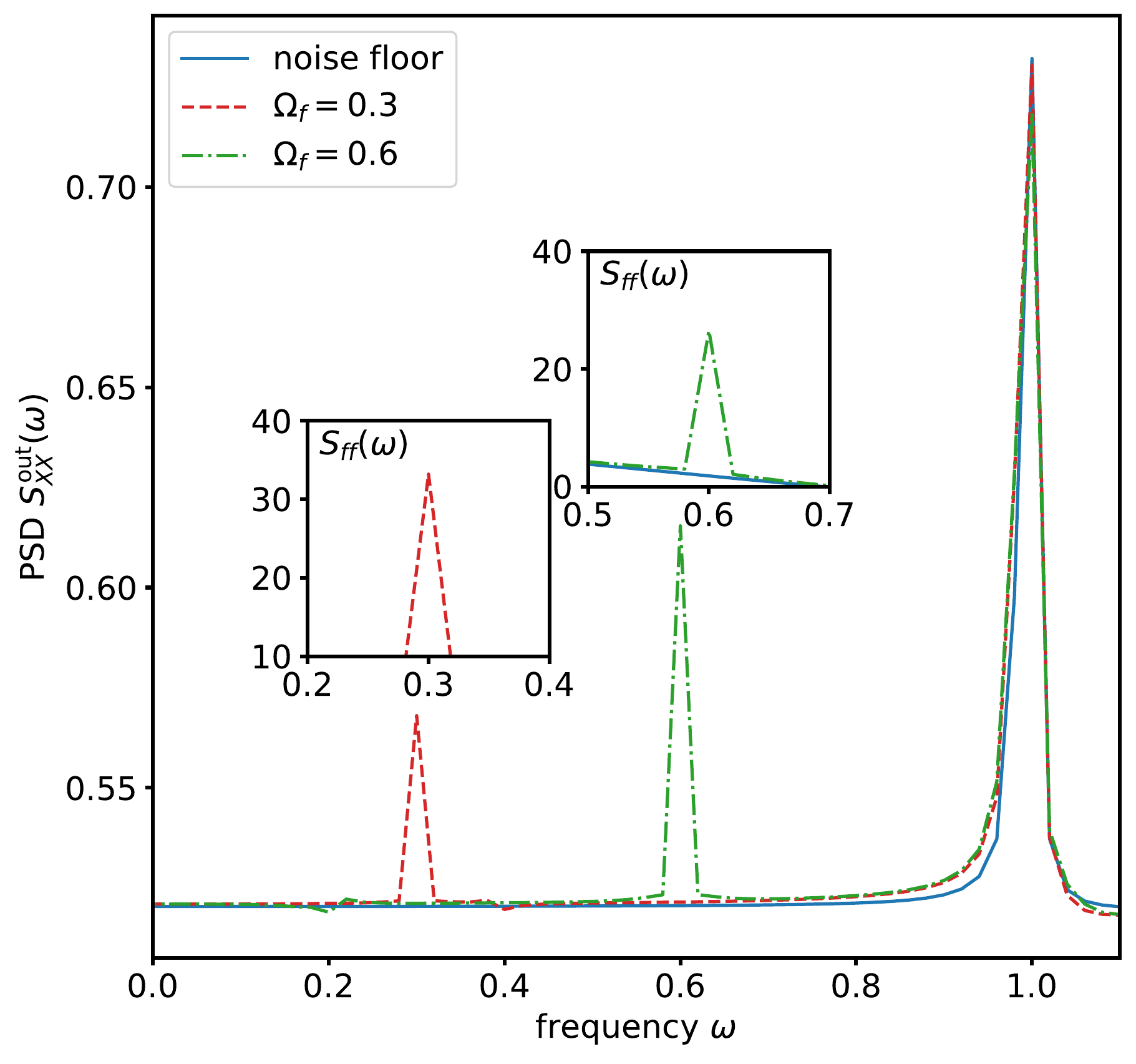}
	\caption{The PSDs of the output optical field in amplitude quadrature $\hat{X}^{\mathrm{out}}$ for different external forces which are obtained from the numerical simulation for a three-microsphere array optomechanical system. The insets show the corresponding PSDs of the forces. The parameters are set as follows: the mechanical frequency $\Omega_{1,2,3} = 1$, the damping rate $\gamma = 0.01$, the thermal phonon number $\bar{n}^{\mathrm{m}}_{1,2,3} = 0.02$, the cavity detuning $\Delta = 5$, the cavity decay rate $\kappa = 5$, the coupling strength $G = 0.1$, and the external force respectively $f_{1,2,3}(t) = 0.1\cos(0.3 t)$ and $f_{1,2,3}(t) = 0.1\cos(0.6 t)$.}
	\label{fig:psd}
\end{figure}

We numerically simulate the dynamics of the whole optomechanical system via the following Markovian Master equation
\begin{equation}\label{master_eq}
	\dot{\hat{\rho}} = \frac{1}{\mathrm{i}\hbar}\left[\hat{\rho},\, \hat{H}_{\mathrm{L}}\right] + \mathcal{L}\left(\hat{\rho}\right),
\end{equation}
where $\hat{\rho}$ is the density operator of the optomechanical system,
$\hat{H}_{\mathrm{L}} = \hbar\Delta_c\hat{a}^\dagger\hat{a}
+ \sum_{i = 1}^{N}\frac{\hbar\Omega}{2}\left(\hat{p}_i^2 + \hat{q}_i^2\right)  + \hbar f_i(t) \hat{q}_i + \hbar G \left(\hat{a}^{\dagger} + \hat{a}\right)\hat{q}_{i}$
is the linearized Hamiltonian, and $f_i(t)$ is the time-dependent external force acting on each mechanical oscillator which induces an acceleration via the relation in Eq. (\ref{eq:force_to_acceleration}), i.e., one can obtain the acceleration through $a_i(t) = \sqrt{\frac{\hbar\Omega_i}{m_i}}f_i(t)$.  
The Lindblad superoperator describing the system dissipation reads
\begin{equation}
	\begin{aligned}
		\mathcal{L}\left(\hat{\rho}\right) =& 
		\frac{\kappa}{2}\left(2\hat{a}\hat{\rho}\hat{a}^{\dagger} - \hat{a}^{\dagger}\hat{a}\hat{\rho} - \hat{\rho}\hat{a}^{\dagger}\hat{a}\right)
		\\
		&+ \sum_{i = 1}^{N}\frac{\gamma}{2}\left(\bar{n}^m_i + 1\right)\left(2\hat{b}_i\hat{\rho}\hat{b}_i^{\dagger} - \hat{b}_i^{\dagger}\hat{b}_i\hat{\rho} - \hat{\rho}\hat{b}_i^{\dagger}\hat{b}_i\right)
		\\
		&+ \frac{\gamma}{2}\bar{n}^{m}_i\left(2\hat{b}_i^{\dagger}\hat{\rho}\hat{b}_i - \hat{b}_i\hat{b}_i^{\dagger}\hat{\rho} - \hat{\rho}\hat{b}_i\hat{b}_i^{\dagger}\right),
	\end{aligned}
\end{equation}
where $\hat{b}_{i}$ is the mechanical annihilation operator of the $i$th sphere.
In the simulation, we assume the external forces acting on each oscillator are identical and in the form of $f(t) = A_f \cos\Omega_{f}t$ with the amplitude $A_f$ and the frequency $\Omega_{f}$. We simulate the dynamical evolution of a three-microsphere array optomechanical system and plot the PSDs of the output optical field in amplitude quadrature $\hat{X}^{\mathrm{out}}$ and the total force $F = \sum_{i=1}^3f_i(t)$ (insets) for different frequencies $\Omega_f$ in Fig. \ref{fig:psd}.
Two peaks at the frequencies corresponding to that of the external force are observed, which confirms that the optomechanical interaction encodes the information of the collective motion induced by the external acceleration into the cavity ﬁeld. We also note that the height of the peaks corresponding to the external force in the optical PSD increases along with the frequency of the acceleration approaching the mechanical resonance frequency due to the Lorentzian-like line shape response functions.
In the insets, the two PSDs of the forces corresponding to the frequencies $\Omega_f = 0.3$ and $\Omega_f = 0.6$ have almost equal height (filled area) because of the same amplitude $A_f = 0.1$.
According to the connection between the amplitude and the PSD in Eq. (\ref{PSD-2}),
we integrate the PSD around the peak $\Omega_f = 0.3$ and $\Omega_f = 0.6$ over the noise floor in the insets and obtain the variance $\langle F^2\rangle|_{\Omega_f = 0.3} \approx 0.043$ 
and $\langle F^2\rangle|_{\Omega_f = 0.6} \approx 0.042$.
This result is well consistent with the variance obtained from the analytical expression $F(t) = 3f(t) = 0.3 \cos\left(\Omega_{f}t\right)$ with $\langle F^2\rangle = 0.045$.

\section{The experimental realization.}

We can employ holographic optical tweezers to levitate many nano-particles inside an optical cavity to realize our acceleration sensing scheme. Following the Ref. \cite{yan2022demand}, a one/two-dimensional array can be created by a pair of orthogonal acousto-optic deflectors (AOD). The resulting beams are focused by an objective lens to create an array of optical tweezers in an optical cavity, which is driven by a laser beam. An auxiliary trapping beam is used to rearrange the pattern of an array and therefore adjust the positions of the nano-particles so that only their collective mode is coupled to the cavity mode. The transmission field of the cavity is measured by a homodyne detection scheme. 

Here, the adjacent distance of two particles (equilibrium positions) should satisfy the condition in Eq. (\ref{eq:position_condition}), which can be adjusted by the integer $n$.
Since the distance between two adjacent optical traps created by holographic optical tweezers is  several microns in experiments, the integer $n$ should be larger than $2$ at least. However, this has no effect on the coupling between the collective mode and cavity mode.

\section{Summary and outlook}

In conclusion, we have proposed a protocol for acceleration sensing by using an optically levitated MSA.
With the help of holographic optical tweezers, $N$ small mass microspheres are trapped and levitated in a driven FP cavity. These small mass spheres form an effective large mass sphere whose mass is distributed in multiple different optical traps, which circumvents the difficulty of levitating a large mass microsphere in a single optical trap. Moreover, the collective motion of the MSA leads to a measurement enhancement effect that can greatly improve acceleration sensitivity. The FP cavity coupled with the MSA is used for the readout of the acceleration information: By suitably adjusting the positions of the microspheres in the cavity, the optomechanical interaction encodes the acceleration information acting on the MSA onto the intracavity photons, which can be measured at the cavity output via the homodyne or heterodyne detection. The results of analytical derivations and numerical simulations show that, compared to the traditional single-microsphere measurement protocol, our scheme presents a $\sqrt{N}$-times improvement in sensitivity.

In the future, along with the development of holographic optical tweezers technique \cite{yang2021optical} in high vacuum and optical measurement methods, our protocol may be extended to a large 3D sensing array which possibly enables full-profile imaging and dynamical evolution monitoring of a force field in real-time. In addition, the optically levitated MSA itself provides a platform to investigate quantum entanglement and quantum correlation \cite{rudolph2022force,rieser2022tunable}.

\appendix*
\section{The derivation of the PSD of acceleration}
Here we describe the detailed derivation of the PSD of the acceleration in Eq. (\ref{eq:PSD_aa}) in the main text.
From the Hamiltonian (\ref{hamiltonian}) and the standard linearization procedure \cite{genes2008ground}, we obtain the quantum Langevin equations QLEs. (\ref{QLEs}) in terms of the quantum fluctuation operators, together with a set of classical Langevin equations (CLEs) in terms of the classical mean values,
\begin{equation}\label{eq:cles}
	\begin{aligned}
		\dot{\alpha} &= -\left(\mathrm{i}\Delta + \frac{\kappa}{2}\right)\alpha  - \mathrm{i}\epsilon,
		\\
		\dot{q}_i^{\mathrm{s}} &= \Omega_i p_i^{\mathrm{s}},
		\\
		\dot{p}_i^{\mathrm{s}} &= - \Omega_i q_i^{\mathrm{s}} - \gamma_i p_i^{\mathrm{s}} - g_i|\alpha|^2.
	\end{aligned}
\end{equation}
By setting the time derivatives to zero in Eqs. (\ref{eq:cles}), one obtains the steady-state mean values as
\begin{equation}
	\begin{aligned}
		\alpha &= \frac{- \mathrm{i}\epsilon}{\mathrm{i}\Delta + \kappa/2}, 
		\\
		\quad p_i^{\mathrm{s}} &= 0,
		\\
		q_i^{\mathrm{s}} &= - \frac{g_i|\alpha|^2}{\Omega_i},
	\end{aligned}
\end{equation}
which are used to determine the effective detuning and coupling strengths in QLEs (\ref{QLEs}).
To solve the QLEs in Eqs. (\ref{eq:qles-2}) in the main text, we transform them into the frequency domain via the Fourier transformation $\hat{\tilde{O}}(\omega) = \int_{-\infty}^{\infty} \hat{O}(t)\exp(\mathrm{i}\omega t)\mathrm{d}t$, where the operator with and without tilde denote the same quantity in time and frequency domain, respectively. The QLEs in the frequency domain reads
\begin{equation}\label{eq:qles_frequency}
	\begin{aligned}
		\left(\frac{\kappa}{2} - \mathrm{i}\omega \right)\hat{\tilde{X}}(\omega) =&  \sqrt{\kappa}\hat{\tilde{X}}^{\mathrm{in}}(\omega) + \Delta\hat{\tilde{Y}}(\omega),
		\\
		\left(\frac{\kappa}{2} - \mathrm{i}\omega \right)\hat{\tilde{Y}}(\omega) =& \sqrt{\kappa}\hat{\tilde{Y}}^{\mathrm{in}}(\omega) - \Delta\hat{\tilde{X}}(\omega) 
		-  \sqrt{2}G\hat{\tilde{Q}}(\omega),
		\\
		\chi_{m}^{-1}(\omega) \hat{\tilde{Q}}(\omega) =& \hat{\tilde{\xi}}(\omega) - \sqrt{2}NG\hat{\tilde{X}}(\omega) - \tilde{F}(\omega),
	\end{aligned}
\end{equation}
where $\chi_{m}(\omega) = \Omega/(\Omega^2 - \omega^2 - \mathrm{i}\gamma\omega)$ is the bare mechanical response function (susceptibility).
By solving Eqs. (\ref{eq:qles_frequency}), we obtain the amplitude and phase quadratures of the cavity field as
\begin{subequations}\label{eq:solution}
	\begin{eqnarray}
		\hat{\tilde{X}}(\omega) &=& H_{XF}(\omega)\tilde{F}(\omega) + H_{X\xi}(\omega)\hat{\tilde{\xi}}(\omega) + H_{XY}(\omega)\hat{\tilde{Y}}^{\mathrm{in}}(\omega) \nonumber
		\\
		&&+ H_{XX}(\omega)\hat{\tilde{X}}^{\mathrm{in}}(\omega),\label{eq:solution_a}
		\\
		\hat{\tilde{Y}}(\omega) &=& H_{YF}(\omega)\tilde{F}(\omega) + H_{Y\xi}(\omega)\hat{\tilde{\xi}}(\omega) + H_{YX}(\omega)\hat{\tilde{X}}^{\mathrm{in}}(\omega) \nonumber
		\\
		&&+ H_{YY}(\omega)\hat{\tilde{Y}}^{\mathrm{in}}(\omega),\label{eq:solution_b}
	\end{eqnarray}    
\end{subequations}
with the corresponding response functions
\begin{subequations}
	\begin{eqnarray}
		&& H_{XF}(\omega) = -H_{X\xi}(\omega) = -\sqrt{2}G\chi_{m}(\omega)\chi(\omega),
		\\
		&&H_{XY}(\omega) = -\sqrt{\kappa}\chi(\omega),
		\\
		&&H_{XX}(\omega) = -\frac{\frac{\kappa}{2} - \mathrm{i}\omega}{\Delta}\sqrt{\kappa}\chi(\omega),
		\\
		&& H_{YF}(\omega) =
		- \frac{\sqrt{2}G\chi_m(\omega)\left(\frac{\kappa}{2} - \mathrm{i}\omega\right)}{2}\chi(\omega),
		\\
		&& H_{Y\xi}(\omega) = - H_{YF}(\omega),
		\\
		&&H_{YX}(\omega) = - \frac{\sqrt{\kappa}\left(\frac{\kappa}{2} - \mathrm{i}\omega\right)^2}{2\Delta}\chi(\omega) - \frac{\sqrt{\kappa}}{\Delta},
		\\
		&&H_{YY}(\omega) = - \frac{\sqrt{\kappa}\left(\frac{\kappa}{2} - \mathrm{i}\omega\right)}{2}\chi(\omega),
	\end{eqnarray}
\end{subequations}
where
\begin{equation}
	\chi(\omega) = \frac{1}{2NG^2\chi_m(\omega) - \Delta - \frac{(\frac{\kappa}{2} - \mathrm{i}\omega)^2}{\Delta}}.
\end{equation}
With the solutions of $\hat{\tilde{X}}(\omega)$ and $\hat{\tilde{Y}}(\omega)$, the PSDs of the cavity field quadratures are given by
\begin{widetext}
	\begin{subequations}
		\begin{eqnarray}
			S_{XX}(\omega) &=& |H_{XF}(\omega)|^2S_{FF}(\omega) + |H_{X\xi}(\omega)|^2S_{\xi\xi}(\omega)
			+ |H_{XY}(\omega)|^2S_{YY}^{\mathrm{in}}(\omega)
			+ |H_{XX}(\omega)|^2S_{XX}^{\mathrm{in}}(\omega),
			\\
			S_{YY}(\omega) &=& |H_{YF}(\omega)|^2S_{FF}(\omega) + |H_{Y\xi}(\omega)|^2S_{\xi\xi}(\omega)
			+ |H_{YX}(\omega)|^2S_{XX}^{\mathrm{in}}(\omega)
			+ |H_{YY}(\omega)|^2S_{YY}^{\mathrm{in}}(\omega),
		\end{eqnarray}
	\end{subequations}
\end{widetext}
from which, we can therefore solve the PSD of the external force in terms of PSD of cavity field in the amplitude or phase quadrature
\begin{widetext}
	\begin{eqnarray}
		S_{FF}(\omega) &=& \frac{S_{XX}(\omega)}{|H_{XF}(\omega)|^2} - S_{\xi\xi}(\omega) - \frac{|H_{XY}(\omega)|^2}{|H_{XF}(\omega)|^2}S_{YY}^{\mathrm{in}}(\omega) 
		- \frac{|H_{XX}(\omega)|^2}{|H_{XF}(\omega)|^2}S_{XX}^{\mathrm{in}}(\omega) \nonumber
		\\
		&=& \frac{S_{YY}(\omega)}{|H_{YF}(\omega)|^2} - S_{\xi\xi}(\omega) - \frac{|H_{YX}(\omega)|^2}{|H_{YF}(\omega)|^2}S_{XX}^{\mathrm{in}}(\omega)
		- \frac{|H_{YY}(\omega)|^2}{|H_{YF}(\omega)|^2}S_{YY}^{\mathrm{in}}(\omega).
	\end{eqnarray}
\end{widetext}

The power spectral densities of the input noise can be calculated via the correlation functions in Eqs. (\ref{eq:optical_correlation}) and (\ref{eq:mechanical_correlation}), and the exact results are given by
\begin{equation}
	\begin{aligned}
		&S_{XX}^{\mathrm{in}}\left(\omega \right) = S_{YY}^{\mathrm{in}}\left(\omega \right) = \frac{1}{2},
		\\
		&S_{\xi_{i}\xi_{i}}\left(\omega \right) = \gamma_i \left(2\bar{n}^{\mathrm{m}}_i + 1\right).
	\end{aligned}
\end{equation}

To establish the connection between $S_{FF}(\omega)$ and quantities measurable directly in experiments, we need to further consult quantum input-output theory. By substituting the input-output relations $\hat{X}^{\mathrm{out}} = \hat{X}^{\mathrm{in}} - \sqrt{\kappa}\hat{X}$ and $\hat{Y}^{\mathrm{out}} = \hat{Y}^{\mathrm{in}} - \sqrt{\kappa}\hat{Y}$ into Eqs. (\ref{eq:solution}), we obtain the amplitude and phase quadratures of the output optical field as
\begin{widetext}
	\begin{eqnarray}
		\hat{\tilde{X}}^{\mathrm{out}}(\omega) &=&
		- \sqrt{\kappa}\left[H_{XF}(\omega)\tilde{F}(\omega) 
		+ H_{X\xi}(\omega)\hat{\tilde{\xi}}(\omega)\right]
		- \sqrt{\kappa}H_{XY}(\omega)\hat{\tilde{Y}}^{\mathrm{in}}(\omega) + \left[1 - \sqrt{\kappa}H_{XX}(\omega)\right]\hat{\tilde{X}}^{\mathrm{in}}(\omega),
		\\
		\hat{\tilde{Y}}^{\mathrm{out}}(\omega) &=&
		- \sqrt{\kappa}\left[H_{YF}(\omega)\tilde{F}(\omega) 
		+ H_{Y\xi}(\omega)\hat{\tilde{\xi}}(\omega)\right] 
		- \sqrt{\kappa}H_{YX}(\omega)\hat{\tilde{X}}^{\mathrm{in}}(\omega)+ \left[1 - \sqrt{\kappa}H_{YY}(\omega)\right]\hat{\tilde{Y}}^{\mathrm{in}}(\omega).		
	\end{eqnarray}
\end{widetext}
By using the relation between the force and acceleration in Eq. (\ref{eq:force_to_acceleration}), one can obtain the PSD of the output optical field in amplitude and phase in terms of the PSD of the acceleration as
\begin{widetext}
	\begin{subequations}
		\begin{eqnarray}
			S_{XX}^{\mathrm{out}}(\omega) =&&
			\frac{N^2 m}{\hbar\Omega}|\sqrt{\kappa} H_{XF}(\omega)|^2 S_{aa}(\omega)
			+ |\sqrt{\kappa} H_{X\xi}(\omega)|^2 S_{\xi\xi}(\omega)
			+ |\sqrt{\kappa}H_{XY}(\omega)|^2S_{YY}^{\mathrm{in}}(\omega)
			\nonumber
			\\&&+ |1 - \sqrt{\kappa}H_{XX}(\omega)|^2S_{XX}^{\mathrm{in}}(\omega),
			\\
			S_{YY}^{\mathrm{out}}(\omega) =&&
			\frac{N^2 m}{\hbar\Omega}|\sqrt{\kappa} H_{YF}(\omega)|^2 S_{aa}(\omega)
			+ |\sqrt{\kappa} H_{Y\xi}(\omega)|^2 S_{\xi\xi}(\omega) + |\sqrt{\kappa}H_{YX}(\omega)|^2S_{XX}^{\mathrm{in}}(\omega)
			\nonumber
			\\&&+ |1 - \sqrt{\kappa}H_{YY}(\omega)|^2S_{YY}^{\mathrm{in}}(\omega).
		\end{eqnarray}
	\end{subequations}
\end{widetext}
By solving equations above, we obtain the eventual expression of PSD of the acceleration determined by the output optical field and the input noises as follows
\begin{widetext}
	\begin{eqnarray}
		S_{aa}(\omega) &=& \frac{\hbar\Omega}{N^2 m}\Bigg\{\frac{S_{XX}^{\mathrm{out}}(\omega)}{|\sqrt{\kappa} H_{XF}(\omega)|^2}
		- S_{\xi\xi}(\omega)
		- \left|\frac{H_{XY}(\omega)}{H_{XF}(\omega)}\right|^2S_{YY}^{\mathrm{in}}(\omega)
		- \left|\frac{1 - \sqrt{\kappa}H_{XX}(\omega)}{\sqrt{\kappa}H_{XF}(\omega)}\right|^2S_{XX}^{\mathrm{in}}(\omega)\Bigg\} \nonumber
		\\
		&=& \frac{\hbar\Omega}{N^2 m}\Bigg\{\frac{S_{YY}^{\mathrm{out}}(\omega)}{|\sqrt{\kappa} H_{YF}(\omega)|^2}
		- S_{\xi\xi}(\omega) 
		- \left|\frac{H_{YX}(\omega)}{H_{YF}(\omega)}\right|^2S_{XX}^{\mathrm{in}}(\omega) 
		- \left|\frac{1 - \sqrt{\kappa}H_{YY}(\omega)}{\sqrt{\kappa}H_{YF}(\omega)}\right|^2S_{YY}^{\mathrm{in}}(\omega)\Bigg\}.
	\end{eqnarray}
\end{widetext}
In experiments, the output optical field PSD $S_{XX}^{\mathrm{out}}(\omega)$ in amplitude quadrature or $S_{YY}^{\mathrm{out}}(\omega)$ in phase quadrature can be obtained directly by the optical homodyne and heterodyne detection.

\begin{acknowledgments}
	We thank Peitong He, Tao Liang, Xiaowen Gao, and Zhenhai Fu for useful discussions. 
	This work is supported by Zhejiang Provincial Natural Science Foundation of China (Grant No. LQ22A040010) and the Major Scientific Research Project of Zhejiang Lab (2019 MB0AD01).
\end{acknowledgments}

\bibliography{ref}

\end{document}